\documentclass{iau} 
\usepackage{graphicx}

\title[LPVs  in Local Group galaxies] 
{AGB population as probes of galaxy structure and evolution}

\author[Atefeh Javadi \& Jacco Th.\,van Loon] 
{Atefeh Javadi$^1$ \and Jacco Th.\,van Loon$^2$}

\affiliation{$^1$School of Astronomy, Institute for Research in Fundamental Sciences
      (IPM), P.O.\ Box 19395-5531, Tehran, Iran \\ email: {\tt atefeh@ipm.ir} \\[\affilskip]
$^2$Lennard-Jones Laboratories, Keele University, ST5 5BG, UK \\email: {\tt j.t.van.loon@keele.ac.uk }}

\pubyear{2018}
\volume{343}
\jname{Why Galaxies Care About AGB Stars:
A Continuing Challenge through Cosmic Time}
\editors{F. Kerschbaum, M. Groenewegen \& H. Olofsson}
\begin{document}

\maketitle

\begin{abstract}
The evolution of galaxies is driven by the birth and death of stars.
AGB stars are at the end points of their evolution
and therefore their luminosities directly reflect their birth mass;
this enables us to reconstruct the star formation history. These
cool stars also produce dust grains that play an important role
in the temperature regulation of the interstellar medium (ISM), chemistry, and the
formation of planets. These stars can be resolved in all of
the nearby galaxies. Therefore, the Local Group of galaxies 
offers us a superb near-field cosmology site. Here we can reconstruct
the formation histories, and probe the structure and dynamics,
of spiral galaxies, of the many dwarf satellite galaxies surrounding
the Milky Way and Andromeda, and of isolated dwarf galaxies. It also 
offers a variety of environments in which to study the detailed processes
of galaxy evolution through studying the mass-loss mechanism and dust
production by cool evolved stars. In this paper, I will first review our
recent efforts to identify mass-losing Asymptotic Giant Branch (AGB) stars and 
red supergiants (RSGs) in Local Group
galaxies and to correlate spatial distributions of the AGB stars of different
mass with galactic structures. Then, I will outline our methodology
to reconstruct the star formation histories using variable pulsating
AGB stars and RSGs and present the results for rates of mass--loss 
and dust production by pulsating AGB stars and their analysis in terms of stellar evolution and galaxy evolution.
\end{abstract}

\firstsection 
\section{Introduction}
Stars can be resolved in all of the Local Group galaxies. This allows
the reconstruction of  star formation histories (SFHs) by modelling
the colour--magnitude diagram, or by using the luminosity distribution
of specific stellar tracers. These galaxies have accurate distances
based on the tip of the red giant branch (RGB), period--luminosity 
relation of relatively young Cepheids, or luminosities of old RR Lyrae.
Cool evolved stars are among the most accessible probes of stellar
populations due to their immense luminosity, from 2000 L$_\odot$ 
for tip--RGB stars, $\sim$10$^4$ L$_\odot$ for asymptotic giant 
branch (AGB) stars, up to a few 10$^5$ L$_\odot$ for red supergiants.
Their spectral energy distributions (SEDs) peak around 1$\mu$m, so
they stand out in the I--band (and reddening is reduced at long
wavelengths). They have low surface gravity causing them to pulsate
radially on timescales of months to years (Yuan \etal\ 2018). The most extreme examples
among these long--period variables (LPVs) are Mira (AGB) variables,
which can reach amplitudes of ten magnitudes at visual wavelengths. 
LPVs vary on timescales (not always strictly periodic) from  $\sim$ 
100 days for low mass AGB stars ($\sim$ 1 M$_\odot$; 10 Gyr old) 
to $\sim$ 1300 days for the dustiest massive AGB stars
($\sim$ 4--8 M$_\odot$; 30--200 Myr old);
$\sim$ 600--900 days for red supergiants ($\sim$ 8--30 M$_\odot$; 10--30 Myr old). 
The variability helps identify these beacons; their luminosities
can be used to reconstruct the star formation history; and
their amplitudes pertain to the process by which they lose matter
and ultimately terminate their evolution (van Loon \etal\ 2008; McDonald \& Zijlstra 2016). 
The diagnostic power of LPVs has been demonstrated in M\,33 
(Javadi \etal\ 2013, 2017) and was illustrated once again
by the discovery of a massive (5 M$_\odot$) LPV in the Sagittarius
dwarf irregular galaxy (Whitelock \etal\ 2017).

In this project we aim: to construct the mass function of
LPVs and derive from this the star formation history in 
different galaxy types; to correlate spatial distributions 
of the LPVs of different mass with galactic structures 
(spheroid, disc and spiral arm components); to measure the rate at
which dust is produced and fed into the ISM; to establish correlations
between the dust production rate, luminosity, and amplitude of  LPVs;
and to compare the {\it in situ} dust replenishment with the amount
of pre--existing dust.

\section{LPVs in nearby galaxies}

Nearby  galaxies  in the Local Group provide excellent opportunities 
for studying dust--producing late stages of stellar
evolution over a wide range of metallicity.  This enables 
to study the detailed processes of galaxy evolution.
Furthermore, we can investigate  the formation histories, and probe
the structure and dynamics, of spiral galaxies, of the many dwarf
satellite galaxies surrounding the Milky Way and Andromeda,
and of isolated dwarf galaxies. 

\subsection{LPVs in M\,33 galaxy}

The only spiral galaxies in the Local Group besides the Milky Way are  M\,31 and M\,33. 
M\,31 is highly inclines (i$\sim$ 77$^\circ$), and extinction therefore remains a problem. 
With inclination of i=56$^\circ$, and at d=950 kpc only slightly more distant, M\,33 
is positioned much more favorably. Also, in contrast to M\,31 the prominent disk of Sc galaxy M\,33
bears evidence of recent star formation.

WFCAM and UIST on the UK InfraRed Telescope (UKIRT) was used  to identify 
mass--losing AGB stars and red supergiants in M\,33 galaxy from the central square 
kpc region to a square degree area (Fig.\ 1). K--band observations were complemented with 
occasional observations in J-- and H--band to provide colour information. 
The photometric catalogue of the disc comprises 403\,734 stars,
among which 4643 stars display large--amplitude 
variability (Javadi \etal\ 2015). Likewise for the bulge we identified 
18\,398 stars among which 812 were identified as exhibiting
large--amplitude variability (Javadi \etal\ 2010).

\begin{figure}[b]
\begin{center}
 \includegraphics[width=5in]{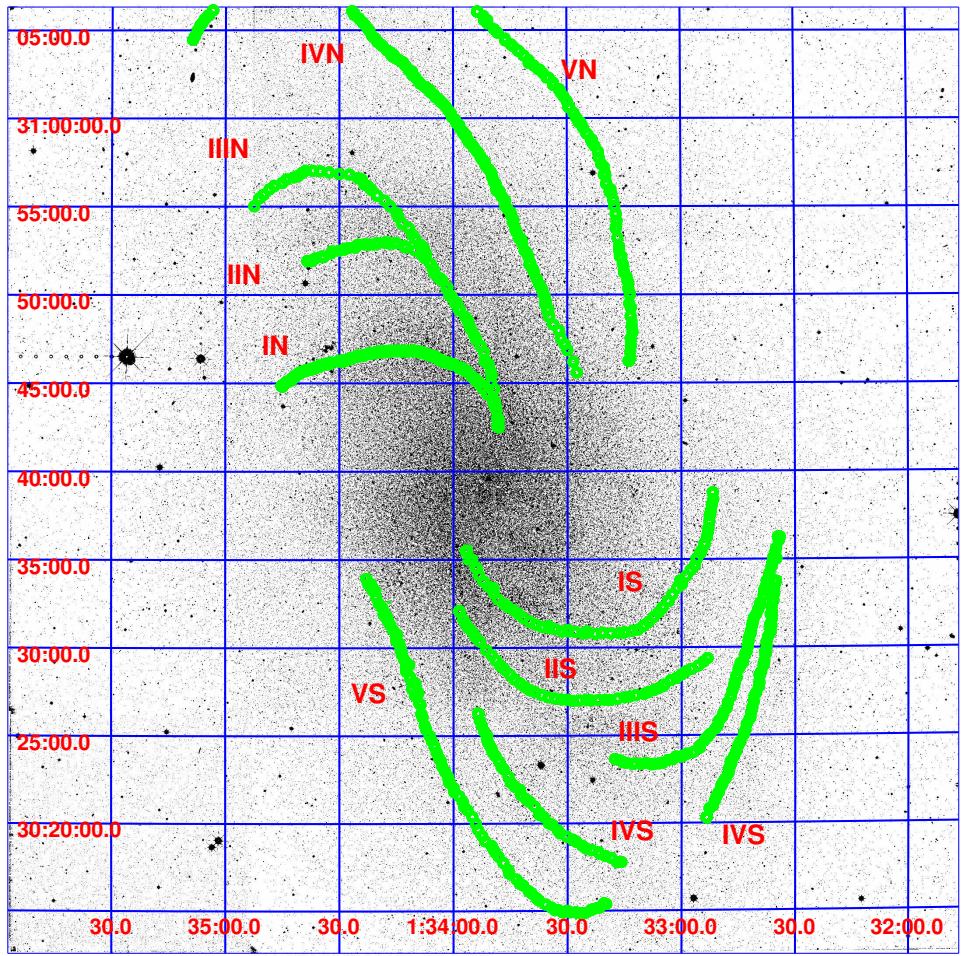} 
 \caption{WFCAM K--band mosaic of M\,33 with the system of
five sets of spiral arms marked on it.}
   \label{fig1}
\end{center}
\end{figure}

\subsection{LPVs in dwarf galaxies}

Among the pertinent questions regarding the dwarf galaxies are to what extent--and when--their
star formation was quenched by gas removal mechanisms, be it as a result of internal feedback (e.g.,\
supernova explosions) or external processes such as the interaction with massive haloes (Weisz \etal\
2015). Can dwarf galaxies be rejuvenated, as some seem to harbour relatively young stars? Is gas only
removed from dwarf galaxies by tidal or ram-pressure stripping, or can it also be (re--)accreted? To what
extent is stellar death able to replenish (metals, dust) the interstellar medium, and to what extent
does it heat it and drive galactic winds? Their star formation histories are among the clearest tracers
of these processes having--or not having--occurred. Understanding the history of dwarf galaxies
may also help us understand stellar streams and minor mergers, and massive globular clusters. 
A description of stellar mass--loss and dust production is of general importance for 
understanding stellar and galaxy evolution. To answer the mentioned questions one
the of robust ways is to identify the LPVs as tracers 
of galaxies star formation histories and chemical enrichment.

The LPVs are identified in some of the galaxies in the Local Group 
via long term monitoring surveys. In addition, a majority of dwarf galaxies have been 
observed with {\it Spitzer} at 3.6 and 4.5 $\mu$m  via DUSTiNGS 
project (DUST in Nearby Galaxies with {\it Spitzer}; Boyer \etal\ 2015a,b).
Due to lack of the monitoring survey 
of dwarf galaxies we started to monitor these galaxies with Isaac Newton Telescope (INT) 
with the purpose of identifying variable AGB stars (Saremi \etal\ 2017).

\subsubsection{INT monitoring survey of dwarf galaxies in the Local Group}

While the Milky Way satellite galaxies are spread all over the sky, a Northern hemisphere survey
alone can be complete for the Andromeda system of satellite galaxies. Such a survey will benefit from
the homogeneity in the distances and hence completeness and accuracy, and foreground populations
and extinction are modest and similar between all Andromeda satellites. Surveying an entire satellite
system enables us to determine variations among satellites due to their infall histories, cosmic
reionization, and internal processes, and to examine how these variations depend on their structural
properties such as total mass, gas mass, and distance to their galaxy host. For instance, the NGC\,147
and NGC\,185 pair are equal in mass but they differ in star formation history and gas content (Weisz
\etal\ 2015). We can also consider the system of satellites as a whole, and add sparse populations
of stellar tracers within individual dwarf galaxies to mimic a much larger galaxy that has sufficient
statistical value. About 20 Andromeda satellites are known to date, and their small number is the
main limit on how clearly one can find trends and variance among them.

Individual Milky Way satellites  observed as a comparison to the Andromeda system
--is the Andromeda system a universal template for galaxy evolution, or just one particular case?
Likewise, isolated galaxies such as Sextans\,A and B-- or the massive and gas--rich IC\,10 serve as
references against which to assess the effects of galactic harassment affecting satellites. While a good
few dwarf galaxies have been monitored over short campaigns to detect RR Lyrae and in some cases
Cepheids, only a few (Southern) galaxies have been monitored (in the infrared) over sufficiently long
periods of time ($>$year) to identify LPVs leaving a vast terrain unexplored. Looking ahead, the most
luminous LPVs can be found as far away as the massive spiral galaxy M\,101 (7 Mpc) to identify dusty
supernova progenitors for spectroscopic follow--up with the James Webb Space Telescope.

We observed in I--band for identification of LPVs, as this is where the contrast between the LPVs
and other stars is greatest, the bolometric corrections to determine luminosities are smallest, and the
effects of attenuation by dust are minimal. However, we also monitored in the V--band.
We prioritized the 62 targets, principally on the basis of their estimated number of AGB stars
--populous galaxies include IC 10 ($>$ 10$^4$ AGB stars) and Sextans A and B ($>$ 10$^3$ AGB stars).
We monitored the entire Andromeda system of satellites; next highest priority was given to isolated
and/or gas--rich galaxies. We included distant globular clusters Pal\,3 and 4, and NGC\,2419 (Galactocentric distances 90--111 kpc) to
investigate their connection to nucleated dwarf galaxies. Ultra--faint Milky Way satellites were given the
lowest priority as they will have few (or no) LPVs. The face--on spiral galaxies M\,101 and NGC\,6946
are included to identify the red supergiant and super--AGB progenitors of imminent supernovae (9 SNe
have been noticed over the past century in NGC\,6946 (The Fireworks Galaxy). The 34$^{\prime}$ wide
field of the INT camera covers each galaxy in one pointing, but dithering between repeat exposures
is required to fill the gaps between the detectors. Even among the Andromeda system of satellites,
none are near enough to one another to fit within one and the same field of view.

In this monitoring survey we aim to [1] reconstruct the SFHs,
[2] perform accurate modelling of their SEDs, and [3] study the relation between pulsation amplitude
and mass--loss (in conjunction with infrared measures of the dust, and theoretical models).

\section{From LPVs counts to SFH}

The LPVs are at the end points of their evolution and therefore their luminosity
can be directly translated into their birth mass; this
enables us to reconstruct the star formation history. 
The star formation history is described by the star formation rate, $\xi$, 
as a function of look-back time (``age''), $t$:
\begin{equation}
\xi(t) = \frac{f(K(M(t)))}{\Delta(M(t))f_{\rm IMF}(M(t))},
\end{equation}
where $f(K)$ is the observed $K$-band distribution of pulsating giant stars,
$\Delta$ is the duration of the evolutionary phase during which these stars
display strong radial pulsation, and $f_{\rm IMF}$ is the Initial Mass
Function describing the relative contribution to star formation by stars of
different mass. Each of these functions depends on the stellar mass, $M$, 
and the mass of a pulsating star at the end of its evolution is directly 
related to its age ($t$) (Fig.\ 2).
This new technique was successfully used in M\,33 (Javadi \etal\ 2011a,b, 2017), 
the Magellanic Clouds (Rezaeikh \etal\ 2014), NGC\,147 \& NGC\,185 (Golshan \etal\ 2017) 
and IC\,1613 (Hashemi \etal\ 2018).  The main results that we found in these 
galaxies are summarized as below:

\begin{itemize}
\item[$\bullet$]{The disc of M\,33 was built $>$ 6 Gyr ago, when 
most stars in M\,33 $\approx73$\% were formed. The second enhanced 
epoch of star formation in M\,33 occurred $\sim250$ Myr ago and contributed 
$\sim6$\% to M\,33's historic star formation. Radial star formation history
profiles suggest that the
inner disc of M 33 was formed in an inside--out formation scenario.}
\item[$\bullet$]{We found a significant difference in the ancient SFH of the 
LMC and the SMC. For the SMC the bulk of the stars formed a few Gyr 
later than the LMC. A secondary peak of SFH at $\sim700$ Myr ago in the LMC and 
the SMC is possibly due to the tidal interaction between the Magellanic Clouds and 
their approach to the Milky Way.}
\item[$\bullet$]{In spite of similar mass and morphological type, NGC\,147 and NGC\, 185,  
which are two of the massive satellites of the Andromeda galaxy reveal 
completely different SFHs. NGC\,185 formed earlier than NGC\,147 but its star 
formation continued until recent times with almost constant rate while the star 
formation in NGC\,147 quenched at least 300 Myr ago.  These results are 
corroborated by strong tidal distortions of NGC\,147 and the presence of gas in the centre 
of NGC\,185.}
\item[$\bullet$]{We do not find any enhanced period of star formation
over the past 5 Gyr in IC\,1613, which suggests that IC\,1613 may have
evolved in isolation for at least that long.}
\end{itemize}

\begin{figure*}
\vbox{\hbox{
\includegraphics[width=2.5in]{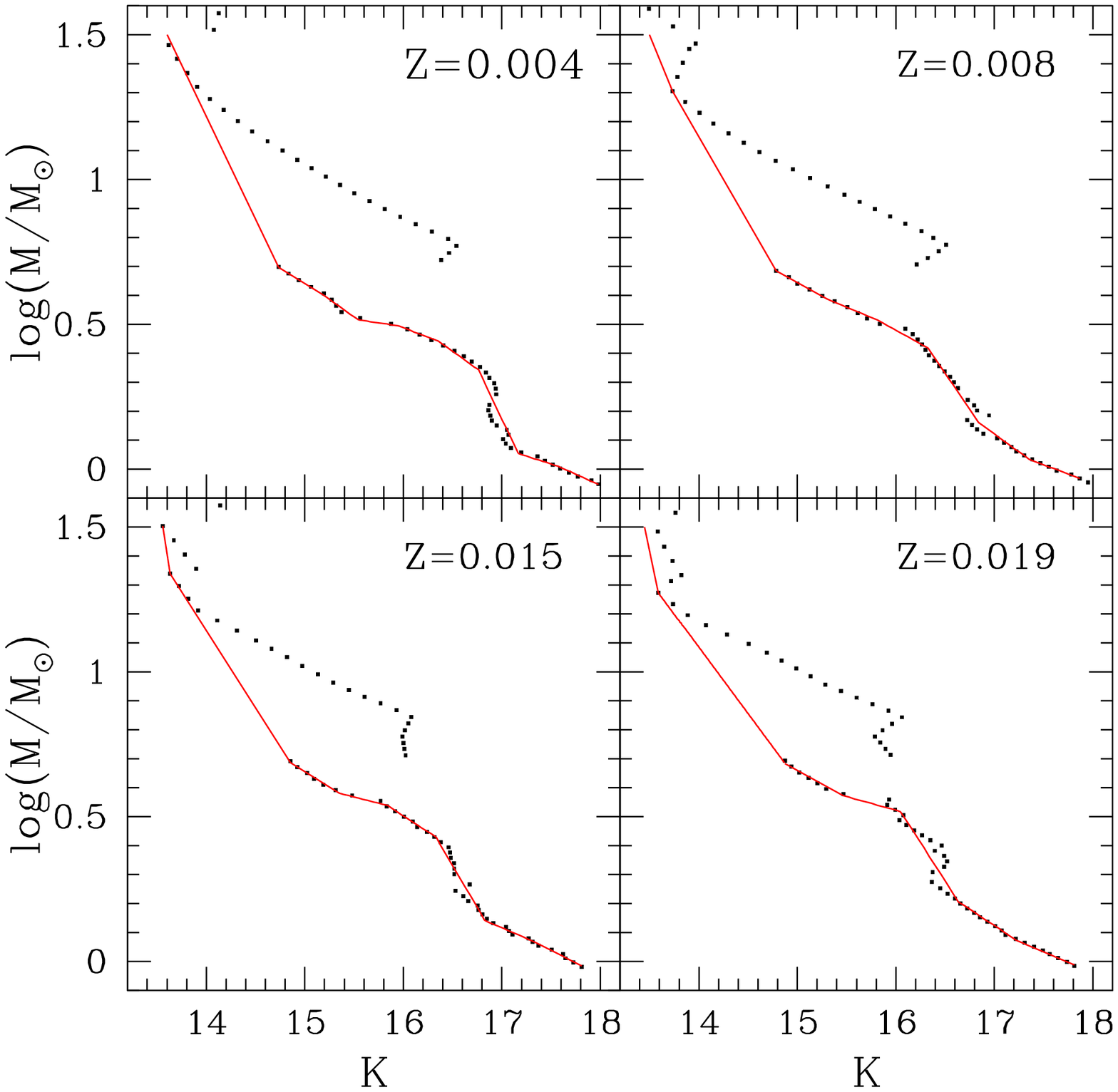} 
\hspace*{1mm}
\includegraphics[width=2.5in]{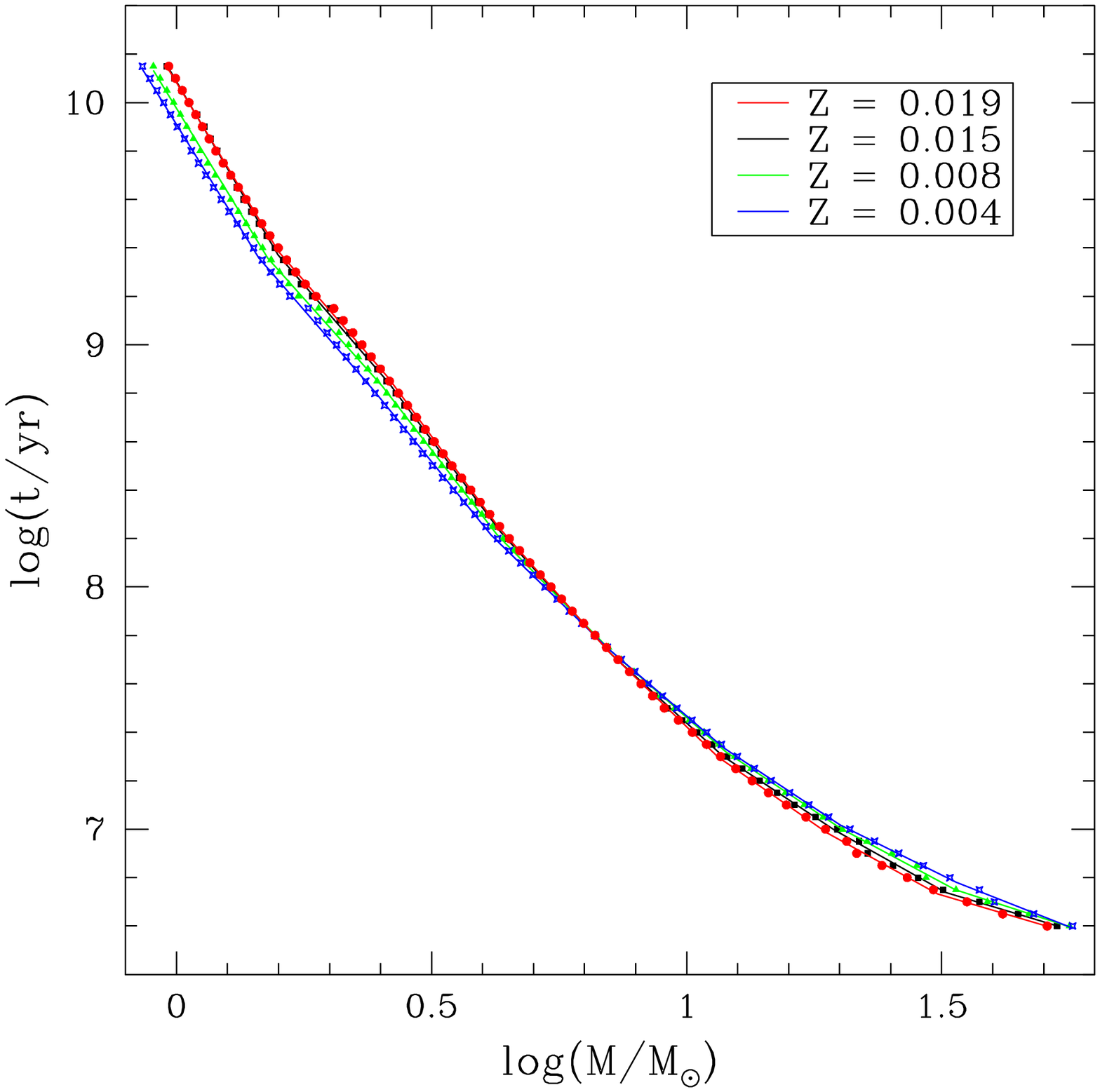} 
}
\vspace*{3mm}
\hbox{
\includegraphics[width=2.5in]{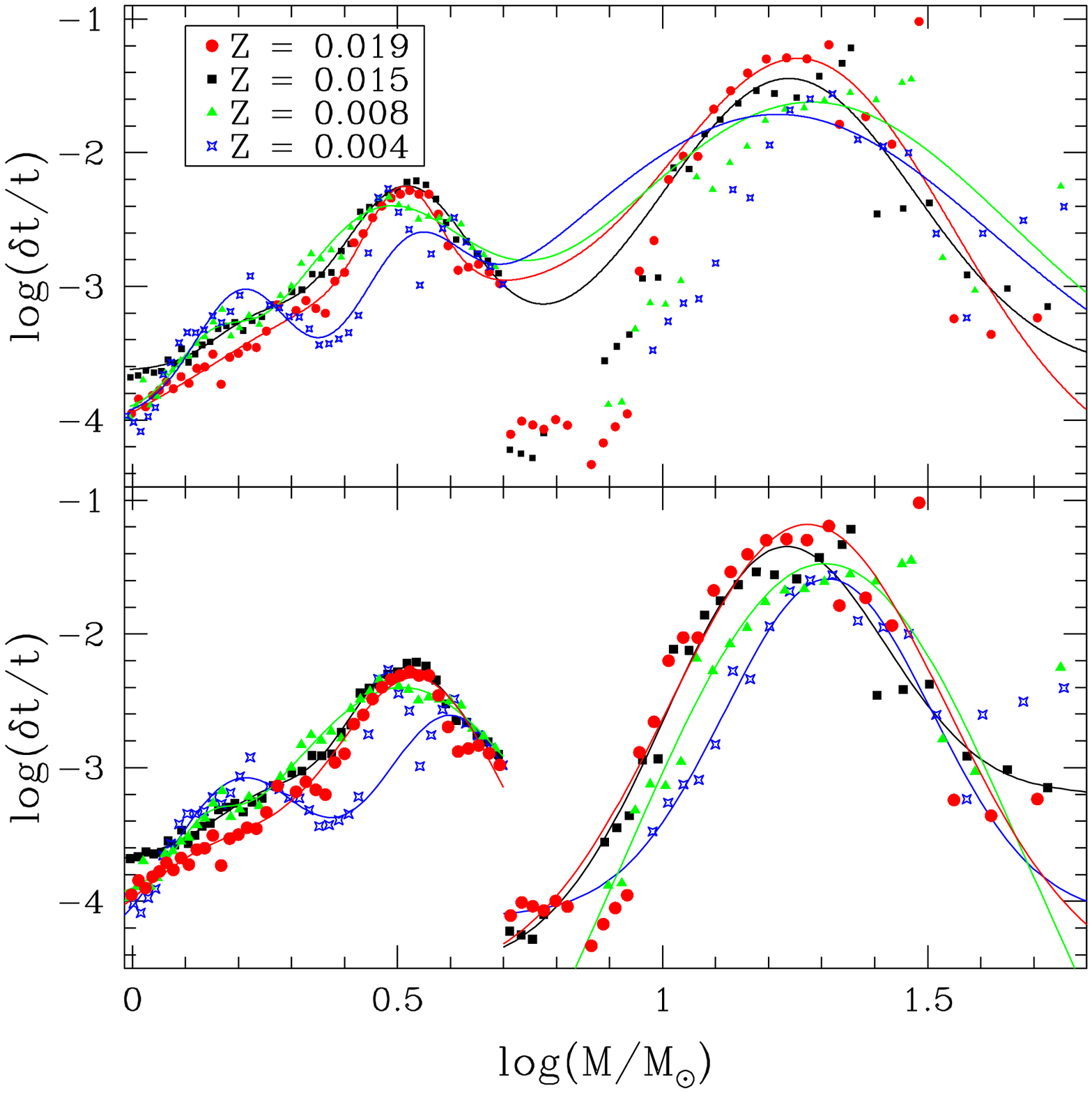} 
\hspace*{1mm}
\includegraphics[width=2.5in]{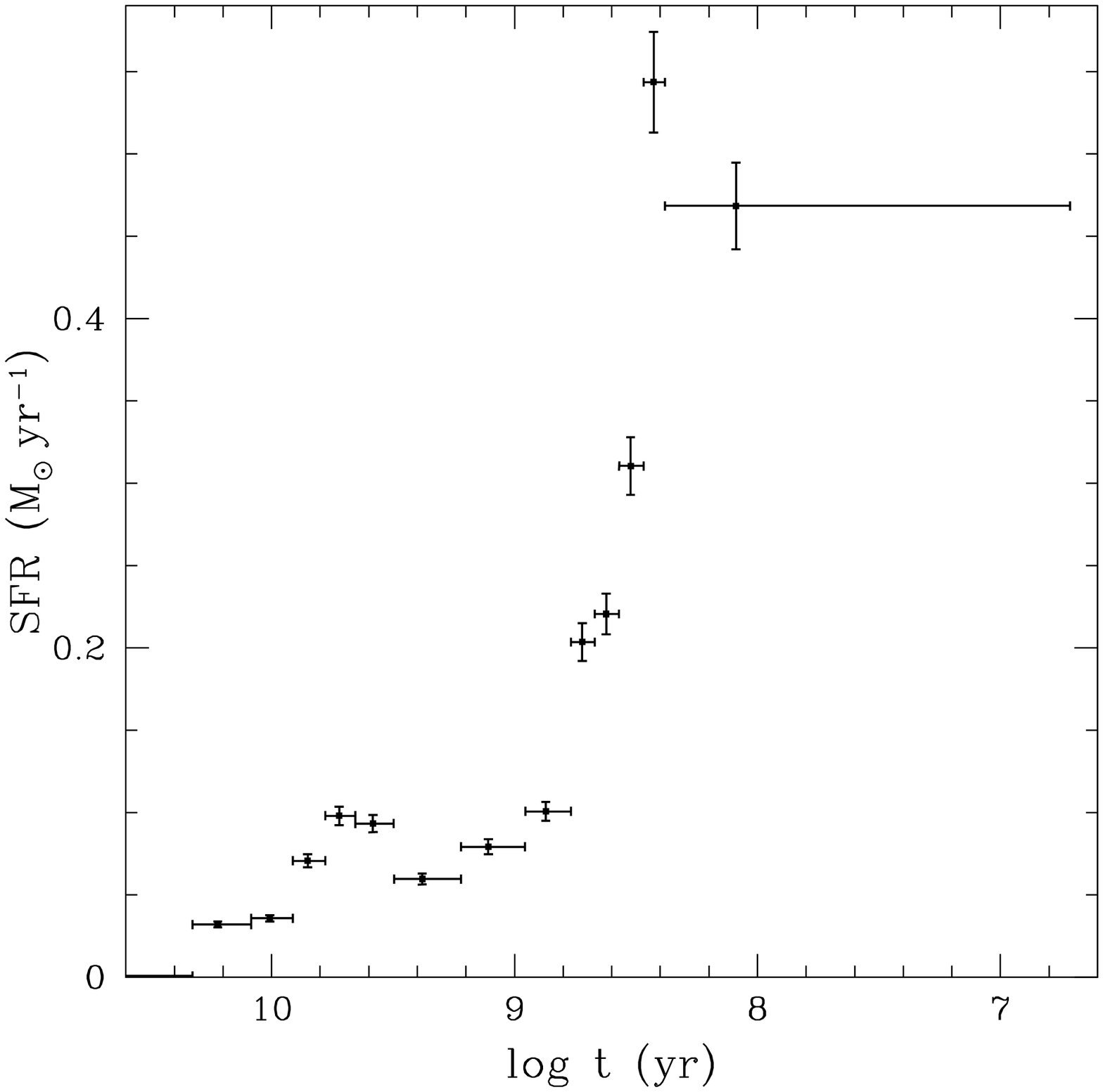} 
}}
\caption[]{From LPVs counts to SFH ({\it Top left:}) 
Mass--Luminosity relation for Z = 0.019, 0.015, 0.008
and 0.004. The solid lines are the linear spline
fits; ({\it Top right:}) (Birth) Mass--Age relation
for AGB stars and red supergiants derived from the
Marigo \etal\ (2008) isochrones; ({\it Bottom
Left:}) Mass--Pulsation relation; ({\it Bottom
right:}) The SFH across the entire disc of M\,33. }
\end{figure*}

\section{AGB stars as probes of galaxy structure}
Populations of stars formed at different times may also 
reveal some of the galactic structures in M\,33 galaxy. To this aim 
we separated the stars in our catalogue into massive stars, AGB stars,
and RGB stars, on the basis of K-band magnitude and J--K colour criteria. 
In the central parts, the AGB distribution shows clear signs of a double-
component profile, with the break occurring around r$\sim0.4$
kpc, so in this case we fitted a S\'ersic profile, with R$_e$ = 0.30 kpc 
and n=1.09 (Javadi \etal\ 2011b). Possibly we are dealing with a bar--like
feature, which is a disc-related structure and may be connected
to the footpoints of the spiral arms.
In addition, the spatial distributions of the massive stars, intermediate-age Asymptotic Giant
Branch (AGB) stars and generally old Red Giant Branch (RGB) stars  in this region suggest that
young and intermediate-age stars were formed within the disc, while the oldest stars
may inhabit a more dynamically-relaxed configuration. Interestingly, the massive stars
concentrate in an area South of the nucleus, and the intermediate--age population shows
signs of a “pseudo--bulge” that however may well be a bar--like feature. 
Furthermore, the distribution of stars with respect to five spiral arms in M\,33, suggests 
that there is no evidence for a lag associated with the density wave 
having passed through the position of evolved stars, or any asymmetry at all. 
This means that spiral arms are transient features  and not part of a global density wave potential.
Based on these results, we concluded that dynamical mixing operates on timescales
$<$ 100 Myr (Javadi \etal\ 2017).
\section{From mass--loss rates of LPVs to chemical enrichment of the galaxies}

The LPVs  are also important sources of dust and gas within galaxies. 
The variability of these cool evolved stars can be used to study their evolutionary
state and mass loss. The pulsations are strongest when the mass loss also becomes strongest--it
is likely that the pulsations help drive the outflow, assisted by dust formation inside the shocked
elevated atmosphere (allowing radiation pressure to drive a wind) and possibly by mechanical or
electro--magnetic waves. To estimate the mass--loss 
rate of variable stars we use the combination of near--IR and mid--IR data. In the 
case of M\,33 almost 2000 variable stars have also been identified by {\it Spitzer}
(Javadi \etal\ 2013, and in preparation). 
We modelled SEDs of variables stars with at least two measurements in near--IR bands and two mid--IR bands 
using the publicly available dust radiative transfer code
{\sc dusty} (Ivezi\'c \& Elitzur 1997). 
In addition 24--$\mu$m sources from Montiel \etal\ (2015) were modelled, because 
they contribute a large fraction to the total dust and mass return (Table 1, Fig.\ 3).  
Our results suggest that 
the mass--loss rate is approximately proportional to luminosity 
(and hence birth mass) with almost weaker dependence 
to pulsation period and/or amplitude (reflecting stellar evolution).
In addition, the total mass lost by evolved stars ($
\dot{M}\sim0.1$ M$_\odot$ yr$^{-1}$) falls short by about
a factor of four to sustain 
stars formation with a current rate, therefore 
requiring external sources of gas supply (Javadi \etal\, in preparation).
\begin{figure}[b]
\begin{center}
 \includegraphics[width=5in]{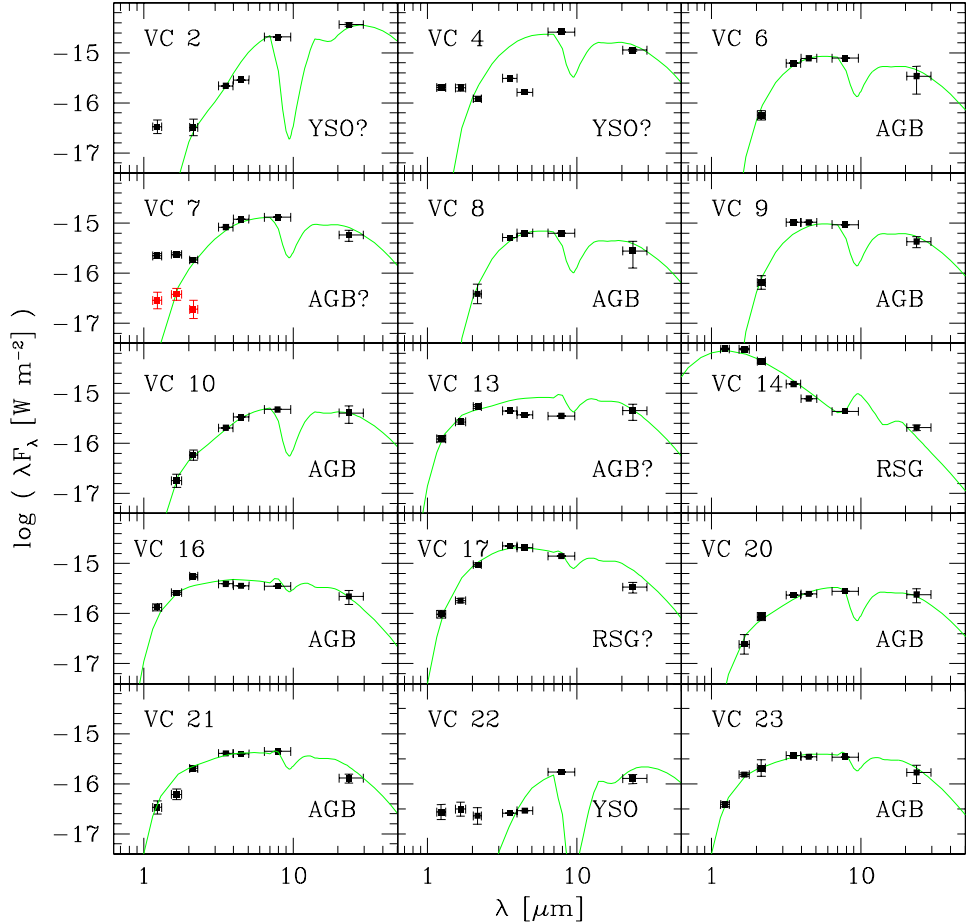} 
 \caption{An example of SEDs for stars with 24--$\mu$m variability.}
   \label{fig1}
\end{center}
\end{figure}

\begin{table}
\begin{center}
\caption{List of 24-$\mu$m variables in M\,33, with UKIRT ID No.\ (Javadi \etal\ 2015).
The luminosities and mass-loss rates are derived from SED fits, which should
be appropriate for AGB stars and RSGs but not YSOs.}
\label{tab1}
{\scriptsize
\begin{tabular}{rrccccll}\hline
name & ID & RA(2000) & DEC (2000) & $\log L/{\rm L}_\odot$ & $\log\dot{M}$ (M$_\odot$ yr$^{-1}$) & variable? & type \\
\hline
 2 & 311369 & 01:34:22.85 & +30:34:09.9 & 5.17 & \llap{$-$}2.79 & no       & YSO? \\
 4 &  39836 & 01:33:32.64 & +30:36:55.5 & 5.06 & \llap{$-$}3.37 & no       & YSO? \\
 6 & 304069 & 01:33:29.70 & +30:24:08.6 & 4.61 & \llap{$-$}3.56 & yes      & AGB  \\
 7 &  17077 & 01:34:12.95 & +30:29:38.5 & 4.81 & \llap{$-$}3.55 & no       & AGB? \\
   & 160486 & 01:34:12.87 & +30:29:40.1 &      &                & no       & AGB? \\
 8 & 305279 & 01:33:28.38 & +30:36:47.9 & 4.52 & \llap{$-$}3.75 & yes      & AGB  \\
 9 & 304597 & 01:34:27.85 & +30:43:40.0 & 4.66 & \llap{$-$}3.65 & yes      & AGB  \\
10 & 252686 & 01:33:50.06 & +30:16:31.7 & 4.41 & \llap{$-$}3.66 & probably & AGB  \\
13 &  16033 & 01:33:26.65 & +30:57:14.4 & 4.76 & \llap{$-$}3.80 & yes      & AGB? \\
14 &    453 & 01:34:12.25 & +30:53:14.1 & 5.46 & \llap{$-$}4.60 & no       & RSG  \\
16 &   8656 & 01:33:47.34 & +30:16:32.4 & 4.52 & \llap{$-$}4.09 & yes      & AGB  \\
17 &  24352 & 01:33:49.86 & +30:52:41.3 & 5.06 & \llap{$-$}3.55 & yes      & RSG? \\
20 & 249854 & 01:33:19.68 & +30:31:05.1 & 4.26 & \llap{$-$}4.04 & probably & AGB  \\
21 &  53418 & 01:33:41.54 & +30:14:12.7 & 4.44 & \llap{$-$}4.03 & yes      & AGB  \\
22 & 182878 & 01:33:37.43 & +30:55:50.4 & 3.96 & \llap{$-$}3.46 & no       & YSO  \\
23 &  43590 & 01:34:09.40 & +30:55:18.2 & 4.41 & \llap{$-$}4.09 & yes      & AGB  \\
\hline
\end{tabular}
}
\end{center}
\end{table}

\section{On--going works and conclusion}

We are currently extending our M\,33 study to the dwarf galaxies in the Local 
Group, to derive star formation history and dust production rate.  We aim to identify all 
LPVs and obtain accurate time--averaged photometry and  amplitudes of variability for 
all red giants and supergiants in the dwarf galaxies at Local Group.

In conclusion, this kind of research which is based on resolved AGB
populations, is very important from both theoretical and observational perspectives; Firstly, it
will give an unprecedented map of the temperature and radius variations as a function of
luminosity and metallicity for mass-losing stars at the end of their evolution, which places
important constraints on stellar evolution models and which is a vital ingredient in the much
sought--after description of the mass--loss process. Secondly, from observational prospective, this
research will gather independent diagnostics of the SFHs of different types of galaxies found
in different environments, which help build a comprehensive picture of galaxy evolution in the
Local Group. These two reasons together show the unprecedented success of AGB stars in
investigating the galaxies formation and evolution.

\section*{Acknowledgments}
AJ would like to thanks from the conference organisers  for support. 
We are grateful for financial support by The Leverhulme Trust
under grant No.\ RF/4/RFG/2007/0297, by the Royal Astronomical Society, and by
the Royal Society under grant No.\ IE130487.

\end{document}